\documentstyle[preprint,tighten,aps,eqsecnum]{revtex}
\begin{document}
\draft
\title{
Three-Nucleon Force\\
and\\
the $\Delta$-Mechanism for Pion Production and Pion Absorption
}

\author{M. T. Pe\~na}
\address{
Centro de Fisica Nuclear da Universidade de Lisboa,
P-1699 Lisboa Codex, Portugal\\
and \\
Continuous Electron Beam Accelerator Facility,
Newport News, Virginia 23606, U.S.A.
}
\author{P. U. Sauer}
\address{
Institut f\"ur Theoretische Physik, Universit\"at Hannover,
D-3000 Hannover 1, Germany
}
\author{A. Stadler}
\address{
Institut f\"ur Theoretische Physik, Universit\"at Hannover,
D-3000 Hannover 1, Germany \\
and\\
Department of Physics,
College of William and Mary, Williamsburg, Virginia 23185,
U.S.A.\cite{ASNow}
}
\author{G. Kortemeyer}
\address{
Institut f\"ur Theoretische Physik, Universit\"at Hannover,
D-3000 Hannover 1, Germany
}
\date{\today}
\maketitle
\begin{abstract}
The description of the three-nucleon system in terms of nucleon
and $\Delta$ degrees of freedom is extended to allow for explicit
pion production (absorption) from single dynamic $\Delta$ de-excitation
(excitation) processes. This mechanism yields an energy dependent
effective three-body
hamiltonean. The Faddeev equations
for the trinucleon bound state are solved with a force model that has
already been tested in the two-nucleon system above pion-production threshold.
The binding energy and other bound state properties are calculated.
The contribution to the effective three-nucleon force arising from the pionic
degrees of freedom is evaluated. The validity of previous coupled-channel
calculations with explicit but stable $\Delta$ isobar components in the
wavefunction is studied.
\end{abstract}
\narrowtext

\section{Introduction}\label{sec1}

Two-nucleon and three-nucleon forces are effective
interactions~\cite{1}.
A truely microscopic description of the nucleus in terms of
quantum chromodynamics with quark and gluon degrees of freedom
would avoid the notion of two-nucleon and three-nucleon forces
altogether. If one views the nucleus in the more traditional
hadronic picture as a system of nucleons, isobars, their
corresponding antiparticles and mesons, one encounters
interactions in form of baryon-meson vertices, but again
no two-nucleon and three-nucleon forces. Two-nucleon
and three-nucleon forces arise when subnucleonic
degrees of freedom are frozen. They are therefore artifacts
of theoreticians who choose to work in a Hilbert space
with a restricted number of degrees of freedom.

At intermediate energies the pion and $\Delta$-isobar degrees
of freedom become active in nuclear reactions. Any efficient
description of nuclear phenomena at intermediate energies has to
treat the pion and $\Delta$-isobar degrees of freedom explicitly
besides the nucleon one. A Hilbert space doing so is illustrated in
Fig.~\ref{fig1}. It contains -- besides the purely nucleonic sector
${\cal H}_N$ -- a sector ${\cal H}_\Delta$ in which
one nucleon is replaced by a $\Delta$-isobar and a sector
${\cal H}_\pi$ in which one pion is added to the nucleons.
That extended, though still rather restricted Hilbert space is
motivated in Ref.~\cite{2}.
A force model acting in that Hilbert space
is developed in Refs.~\cite{2} and \cite{3}.
Its hamiltonian is diagrammatically
defined in Fig.~\ref{fig2}. It is not covariant.
Its unlinked one-baryon
processes (e) and (f) can fully account for pion-nucleon scattering
up to 300 MeV pion lab energy. Those processes are to be
parametrized consistently with data.
The $\Delta$-isobar of the force model is a bare particle which only
by its coupling to pion-nucleon states describes the physical
$P_{33}$ resonance. The force model builds up the mechanism for
pion production or pion absorption as a two-step process,
i.e.,
\begin{itemize}
\item by the excitation of a nucleon to or by the deexcitation
of a $\Delta$-isobar through the instantaneous transition potential (b)
and
\item by the subsequent decay of that $\Delta$-isobar into or by its
formation from pion-nucleon states through the pion-nucleon-$\Delta$ vertex
(e).
\end{itemize}
The force model is tested in the
two-nucleon system above pion-threshold, e.g., in Refs.~\cite{4} and
\cite{5} for
all reactions with at most one pion, coupled by unitarity.
It is not tuned yet and therefore fails to account for many observables
as any nucleon-nucleon potential would without a proper fit. However,
the force model is constructed to account for nucleon-nucleon
scattering below pion-threshold with satisfactory quality. A comparison
with other force models of similar structure is given in Ref.~\cite{6}.

The force model of Fig.~\ref{fig2} is not only
meant to account for the two-nucleon
system below and above pion threshold. In heavier nuclei it can also
provide a microscopic basis for describing reactions at intermediate energies,
and the force model has been employed that way \cite{7}. When applied to
nuclear structure problems, the force model yields corrections \cite{8}
for the picture of the nucleus as a system of nucleons only interacting through
two-nucleon forces with each other and through single-nucleon currents with
external probes. E.g., when applied to the trinucleon bound state,
it yields corrections to the effective three-nucleon force and to the effective
two- and three-nucleon currents \cite{9,10}. Characteristic examples for
contributions to the three-nucleon force are illustrated in Fig.~\ref{fig3}.
It has been argued \cite{11,12} and shown \cite{13} that the contributions
arising from the $\Delta$-isobar and the pion degrees of
freedom are most important in a full and realistic three-nucleon force.
Subject of this paper is the relation between the three-nucleon force
and the force model of Fig.~\ref{fig2} with $\Delta$-isobar
and pion degrees of freedom. Preliminary discussions of that theme have
been given in Refs.~\cite{14} and \cite{15}.

Sect.~\ref{sec2} recalls the precise definition of the force model of
Fig.~\ref{fig2}.
It applies the force model to the trinucleon bound state. The resulting
set of equations is very close to those of a coupled-channel treatment
\cite{9,10} for nuclear bound states, in which the $\Delta$-isobar
is considered a stable particle with fixed mass. Sect.~\ref{sec3} describes
the actual calculations carried out in this paper for the full force model
and for approximated variants of it. The calculations are meant to explore
the validity of the corresponding coupled-channel
description of nuclear bound states. Sect.~\ref{sec4} presents the
obtained results. Sect.~\ref{sec5} discusses conclusions.

\section{Application of the Force Model
with $\Delta$-Isobar and Pion Degrees of Freedom
to the Trinucleon Bound State}\label{sec2}

The hamiltonian $H$ of the force model acts in the three sectors of
Hilbert space and couples them. The projectors on the two baryonic
sectors ${\cal H}_N$ and ${\cal H}_\Delta$ are denoted by
$P_N$ and $P_\Delta$, respectively, with the abbreviation $P=P_N+P_\Delta$,
the projector on the sector ${\cal H}_\pi$ with a pion by $Q$. Thus,
$P_N+P_\Delta+Q=1$. The kinetic part $H_0$ of the hamiltonian defines the
Hilbert sectors and commutes with the projectors. It consists of
the individual baryonic contributions $h_0(i)$, $i$ being the baryon
label; in the Hilbert sector ${\cal H}_\pi$ it has the additional pion
contribution $h_0(\pi)$. It includes
rest masses. The mass $m_\Delta^0$ of the bare $\Delta$-isobar used in the
definitions of the Hilbert sector ${\cal H}_\Delta$ and of the force model
is unobservable. The single-particle momentum of a nucleon, a $\Delta$-isobar
and a pion is ${\bf k}_N$, ${\bf k}_\Delta$ and ${\bf k}_\pi$, respectively.
The kinetic energies of the nucleon and of the
$\Delta$-isobar are taken to be nonrelativistic, i.e., $\varepsilon_N(k_N) =
m_N+k_N^2/2m_N$ and $\varepsilon_\Delta(k_\Delta)=m_\Delta^0+
k_\Delta^2/2m_\Delta^0$, the kinetic energy of
the pion to be relativistic, i.e., $\omega_\pi(k_\pi) = \sqrt{m_\pi^2 +
k_\pi^2}$. The interaction part $H_1$ of the
hamiltonian is built from instantaneous unretarded potentials.
It connects the pionic sector only to the one with a $\Delta$-isobar,
thus, $P_NH_1Q=QH_1P_N=0$. The interaction part of the
hamiltonian takes the form
\begin{eqnarray}\label{2.1}
\nonumber
H_1&=&(P_N+P_\Delta)H_1(P_N+P_\Delta)+\\
&&P_\Delta H_1Q+Q H_1P_\Delta+ QH_1Q\ .
\end{eqnarray}

The Schr\"odinger equation for the trinucleon bound-state energy
$E_B$ and for the wave function $|\Psi_B\rangle $
\begin{equation}\label{2.2}
[H_0+H_1]|\Psi_B\rangle =E_B|\Psi_B\rangle
\end{equation}
is projected onto the baryonic and pionic sectors of Hilbert space,
i.e.,
\begin{mathletters}
\[[P(H_0+\delta H_0(E_B))P+P(H_1+\delta H_1(E_B))P]P|\Psi_B\rangle \]
\begin{equation}\label{2.3a}
\qquad\quad=E_BP|\Psi_B\rangle ,
\end{equation}
\begin{equation}\label{2.3b}
Q|\Psi_B\rangle =\frac Q{E_B-QHQ}QH_1P\ P|\Psi_B\rangle ,
\end{equation}
\begin{equation}\label{2.3c}
\langle \Psi_B|P|\Psi_B\rangle +\langle \Psi_B|Q|\Psi_B\rangle =1\ .
\end{equation}
\end{mathletters}
The triton binding energy $E_T$ without rest masses, i.e., $E_T=E_B-3m_N$,
and the baryonic components
$(P_N+P_\Delta)|\Psi_B\rangle $ of the trinucleon wave function follow from
solving the projected equation (\ref{2.3a}).
The pionic component $Q|\Psi_B\rangle $ of the wave function is obtained from
its $\Delta$-component $P_\Delta|\Psi_B\rangle $
according to Eq.~(\ref{2.3b}) by quadrature.
The projected equation (\ref{2.3a}) contains the energy-dependent parts
$\delta H_0(E_B)$ and $\delta H_1(E_B)$, i.e.,
\begin{mathletters}
\begin{eqnarray}
\label{2.4a}
P\delta H_0(z)P&=&\left[PH_1Q\frac Q{z-QHQ}QH_1P
\right]_{\mbox{\tiny disconnected}}\ ,\\
\label{2.4b}
P\delta H_1(z)P&=&\left[PH_1Q\frac Q{z-QHQ}QH_1P
\right]_{\mbox{\tiny connected}}\ .
\end{eqnarray}
\end{mathletters}
The disconnected part $\delta H_0(z)$ is of one-baryon nature, the
connected part $\delta H_1(z)$ contains two-baryon {\it and
possibly} three-baryon pieces according to Fig.~\ref{fig4}. Both parts
are only
defined in the baryonic sectors of Hilbert space and are -- in the
considered force model -- nonzero only
in the one with a $\Delta$-isobar, i.e., $P\delta H_0(z)P_N=P_N\delta
H_0(z)P=P\delta H_1(z) P_N=P_N\delta H_1(z)P=0$. They show an
energy dependence, though the original hamiltonian acts
instantaneously without time delay. That energy dependence
arises from projecting the
pionic component out from the wave function. However, by that energy
dependence the pionic
component preserves its active presence.
Thus, the energy dependence in $\delta H_0(z)$ and $\delta H_1(z)$ is
{\it necessary} and it is well regulated, always prescribed without any
arbitrariness in all applications. This paper only deals with the
trinucleon bound state; but clearly the same set of equations
(2.3)
hold for scattering problems: Due to the energy dependence
of the effective baryonic hamiltonian
$P(H_0+\delta H_0(z)+H_1+\delta H_1(z))P$ the
wave function components of bound and scattering states,
projected onto the baryonic sectors, are
not orthogonal, though they belong to states of different
three-nucleon energy. However, the formalism naturally restores
the orthogonality for the full states which include their
pionic components. In the same way only the full bound-state wave
function is to be normalized according to Eq.~(\ref{2.3c}). The controlled
energy dependence of the effective baryonic hamiltonian is in contrast to a
phenomenologically chosen energy dependence of the two-nucleon potential
as used once in a while\cite{20};
in the latter case there are no rules which affect
the change in the energy dependence to be adopted for different
applications.

Despite the energy dependence of the effective hamiltonian
$P(H_0+\delta H_0(z)+H_1+\delta H_1(z))P$ the projected baryonic equation
(\ref{2.3a}) can be decomposed into a set of equations for Faddeev
amplitudes in the standard way. We introduce the notation
\begin{equation}\label{2.5}
G_0^P(z)=\frac P{z-P[H_0+\delta H_0(z)]P}
\end{equation}
for the effective resolvent and
\begin{equation}\label{2.6}
P[H_1+\delta H_1(z)]P=\sum_iv_i(z)+\sum_iW_i(z)
\end{equation}
for the effective baryonic interaction with $v_i(z)$ denoting the
two-baryon interaction between the pair $(jk)$, $(ijk)$ cyclic,
and with $W_i(z)$ denoting the three-baryon interaction,
particle $i$ being the $\Delta$-isobar in the force before
interaction. The effective two-baryon
interaction $v_i(z)$ has instantaneous contributions arising from
$H_1$ in and between both baryonic sectors and energy-dependent
contributions arising from $\delta H_1(z)$
in the Hilbert sector ${\cal H}_\Delta$ with a $\Delta$-isobar.
The effective three-baryon interaction
$W_i(z)$ of the discussed force model only has energy-dependent
contributions arising from $\delta H_1(z)$ and consequently it is
nonzero only in the baryonic sector ${\cal H}_\Delta$.
Process (e) of Fig.~\ref{fig4} depends on the coordinates of all three baryons,
all three of them interact,
it therefore yields a true three-baryon force $W_i(z)$. However, it is a
singular one through the $\delta$-function for the $\Delta$-isobar momentum;
in fact, it is unlinked, though it is derived from Eq.~(2.4b) and labelled
there otherwise.
Thus, it has the mathematical structure of a two-body interaction $v_i(z)$ in
a three-particle Hilbert space. The Appendix \ref{appA} derives the
mathematical structure of that
particular contribution and shows how it is to be
combined with the two-baryon interactions
in the calculational treatment.

Using the two-baryon interaction $v_i(z)$ for defining the transition matrix
\begin{equation}\label{2.7}
T_i(z)=v_i(z)\left[1+G_0^P(z)\right]T_i(z)
\end{equation}
in the three-baryon space and introducing baryonic Faddeev amplitudes
$P|\psi_i\rangle $, i.e.,
\begin{equation}\label{2.8}
P|\psi_i\rangle =G_0^P(E_B)\left[v_i(E_B)+W_i(E_B)\right]P|\Psi_B\rangle \ ,
\end{equation}
the effective Schr\"odinger equation (2.3a)
gets equivalent to the set
of Faddeev equations
\begin{mathletters}
\begin{eqnarray}
\label{2.9a}
P|\psi_i\rangle &=&G_0^P(E_B)\{T_i(E_B)[P_{ijk}+P_{ikj}] \nonumber\\
&&\ +[1+T_i(E_B)G_0^P(E_B)]W_i(E_B) \nonumber\\
&&\ \times [1+P_{ijk}+P_{ikj}]\}P|\psi_i\rangle \ ,\\
\label{2.9b}
P|\Psi_B\rangle &=&N[1+P_{ijk}+P_{ikj}]P|\psi_i\rangle \ .
\end{eqnarray}
\end{mathletters}
The particular form (\ref{2.9a}) of Faddeev equations is introduced in
Ref.~\cite{16} for the case in which explicit and irreducible
three-body forces are
present. The Faddeev equations (\ref{2.9a}) greatly simplify, once
three-body forces are absent. In Eqs.~(2.9)
$P_{ijk}$ and
$P_{ikj}$ are cyclic and anticyclic permutation operators of
three particles $(ijk)$. In Eq.~(\ref{2.9b}) $N$ is a
normalization constant which only the normalization condition
(\ref{2.3c}) for the full wave function determines.
\section{Calculational Apparatus}\label{sec3}
Calculations of the trinucleon bound state are carried out for the force model
of Fig.~\ref{fig2} and for variants of it. It has been tested in the
two-nucleon system above pion threshold by Refs.~\cite{4} and \cite{5}.
The force model is used there
and for this paper in the approximation $QH_1Q=0$, which neglets
all interactions in the pionic sector ${\cal H}_\pi$ of the Hilbert space.

Since $QH_1Q=0$,
all energy-dependent three-baryon contributions to the effective
baryonic interaction (\ref{2.6}) disappear and the Faddeev equations
(\ref{2.9a}) simplify to those with two-baryon forces only, i.e.,
\begin{equation}\label{3.1}
P|\psi_i\rangle =G_0^P(E_B)T_i(E_B)[P_{ijk}+P_{ikj}]P|\psi_i\rangle \ .
\end{equation}
The only energy-dependent contributions to the effective baryonic
hamiltonian (\ref{2.3a}) which survive are the processes (a) and (b) of
Fig.~\ref{fig4}. Both are determined by the pion-nucleon-$\Delta$
vertex $QH_1P_\Delta$ which is calibrated through pion-nucleon
scattering in the $P_{33}$ partial wave.
The single-baryon nature of the vertex is made explicit by the notation
$QH_1P_\Delta=\sum_iQh_1(i)P_\Delta$, $i$ being the label of the baryon which
is transformed from a $\Delta$-isobar to a pion-nucleon state.
Process (a) yields the $\Delta$-isobar
self-energy correction $P_\Delta\delta H_0(z)P_\Delta$ in the three-baryon
resolvent, i.e.,
\widetext
\begin{mathletters}
\begin{equation}\label{3.2a}
P_\Delta\delta H_0(z)P_\Delta
=\sum_iP_\Delta h_1(i)Q\frac Q{[z-Qh_0(j)Q-Qh_0(k)Q]-Qh_0(i)Q-
Qh_0(\pi)Q}Qh_1(i)P_\Delta\ ,
\end{equation}
\narrowtext\noindent
process (b) the retarded one-pi\-on ex\-change
$P_\Delta\delta H_1(z)P_\Delta$ in
the effective two-baryon interaction, i.e.,
\widetext
\begin{equation}\label{3.2b}
P_\Delta\delta H_1(z)P_\Delta
=\sum_{j\neq k}P_\Delta h_1(k)Q\frac Q{[z-Qh_0(i)Q]-Qh_0(j)Q-Qh_0(k)Q-
Qh_0(\pi)Q}Qh_1(j)P_\Delta\ .
\end{equation}
\end{mathletters}
\narrowtext
In the single-baryon part $P_\Delta\delta H_0(z)P_\Delta$  the
operator $[z-Qh_0(j)Q-Qh_0(k)Q]$ of the three-baryon
resolvent reflects the fact that
at the available energy $z$ the non\-in\-ter\-acting nucleons $j$ and $k$ are
present and propagate besides the $\Delta$-isobar $i$; that operator becomes
a $c$-number parameter in a three-baryon momentum-space basis. In the
two-baryon part $P_\Delta\delta H_1(z)P_\Delta$ the operator
$[z-Qh_0(i)Q]$ of the resolvent reflects the fact that at the available
energy $z$ the noninteracting nucleon $i$ is present and propagates beside
the interacting nucleon-$\Delta$ pair $(jk)$; that operator becomes a
$c$-number parameter in a three-baryon momentum-space basis. Both
retarded contributions $P_\Delta\delta H_0(z)P_\Delta$ and
$P_\Delta\delta H_1(z)P_\Delta$ are defined in Eqs.~(3.2)
for the three-baryon
system. However, they notice the presence and absence of noninteracting
particles through their energy dependence, they are therefore different,
e.g., in one-baryon and two-baryon systems.
The structure of both parts $P_\Delta\delta H_0(z)P_\Delta$ and
$P_\Delta\delta H_1(z)P_\Delta$ and their use in the trinucleon
bound-state calculation are now discussed.
\subsection{Three-Baryon Basis States}\label{sec3.1}
The three-baryon basis states $|p_1q_1\nu_1\rangle$,
required for the calculation in the Hilbert sectors
${\cal H}_N$ and ${\cal H}_\Delta$ are diagramatically defined in
Fig.~\ref{fig5}, $(p_1q_1)$ are the
magnitudes of the Jacobi momenta and $\nu$ abbreviates all discrete
quantum numbers. The calculation is done in the trinucleon c.m. system,
thus, the total momentum ${\bf k}$ is zero. In the purely nucleonic Hilbert
sector ${\cal H}_N$ the definition of the Jacobi momenta is standard.
In the Hilbert sector ${\cal H}_\Delta$ the Jacobi momenta
are nonrelativistically
defined with the mass $m_N$ for nucleons and $m_\Delta^0$ for the
$\Delta$-isobar. The momenta $(p_1q_1)$ are denoted by $(p_\Delta q_\Delta)$
when the $\Delta$-isobar is the spectator particle 1, i.e.,
\begin{mathletters}
\begin{eqnarray}\label{3.1a}
{\bf p}_\Delta&=&\frac{m_N{\bf k}_{N2}-m_N{\bf k}_{N3}}{2m_N}\ ,\\
{\bf q}_\Delta&=&\frac{m_\Delta^0({\bf k}_{N2}+{\bf k}_{N3})-2m_N{\bf
k}_\Delta}
{m_\Delta^0+2m_N}\ ,\\
{\bf k}&=&{\bf k}_\Delta+{\bf k}_{N2}+{\bf k}_{N3}=0\ ,
\end{eqnarray}
\end{mathletters}
\begin{eqnarray}
&&H_0|p_\Delta q_\Delta\nu_\Delta\rangle_1= \nonumber\\
&&\left[2m_N+\frac{p_\Delta^2}{m_N}+
\frac{q_\Delta^2}{4m_N}+m_\Delta^0+\frac{q_\Delta^2}{2m_\Delta^0}\right]
|p_\Delta q_\Delta\nu_\Delta\rangle_1\label{3.2}
\end{eqnarray}
with ${\bf k}_\Delta,\ {\bf k}_{N2}$ and ${\bf k}_{N3}$ being single-particle
momenta of the baryons. The momenta $(p_1q_1)$ are denoted by $(p_Nq_N)$
when a nucleon is the spectator particle 1, i.e.,
\begin{mathletters}
\begin{eqnarray}
{\bf p}_N&=&\frac{m_\Delta^0{\bf k}_{N2}-m_N{\bf k}_\Delta}{m_N+m_\Delta^0}\
,\\
{\bf q}_N&=&\frac{m_N({\bf k}_{N2}+{\bf k}_\Delta)-(m_N+m_\Delta^0){\bf
k}_{N1}}
{2m_N+m_\Delta^0}\ ,\\
{\bf k}&=&{\bf k}_{N1}+{\bf k}_{N2}+{\bf k}_\Delta=0\ ,
\end{eqnarray}
\end{mathletters}
\begin{eqnarray}\label{3.4}
&&H_0|p_N q_N\nu_N\rangle_1= \nonumber\\
&&\left[m_N+m_\Delta^0+\frac{p_N^2}2\left(\frac1{m_N}+\frac1{m_\Delta^0}
\right)+
\frac{q_N^2}{2(m_N+m_\Delta^0)}\right. \nonumber\\
&&\quad\left.+m_N+\frac{q_N^2}{2m_N}\right]
|p_N q_N\nu_N\rangle_1
\end{eqnarray}
The basis states are antisymmetrized
with respect to particles 2 and 3.
In the latter case $|p_Nq_N\nu_N\rangle_1$ the
$\Delta$-isobar is taken to be particle 3 before antisymmetrization.
The basis states $|p_\Delta q_\Delta\nu_\Delta\rangle_1$ and
$|p_N q_N\nu_N\rangle_1$
are orthogonal to each other; they are different states in the complete
set of states describing two nucleons and one $\Delta$-isobar.

\subsection{The $\Delta$-Isobar Self-Energy Correction
$P_\Delta\delta H_0(z)P_\Delta$ and
the Effective Three-Baryon Resolvent}\label{sec3.2}
The effective three-baryon resolvent (\ref{2.5}) is illustrated in
Fig.~\ref{fig6}. It is trivial in the Hilbert sector ${\cal H}_N$, but
receives the pionic correction $P_\Delta\delta H_0(z)P_\Delta$ in the
sector ${\cal H}_\Delta$ with one $\Delta$-isobar. That pionic correction is
also seen in $P_{33}$ pion-nucleon scattering as illustrated in
Fig.~\ref{fig7}, and it is calibrated there.

The pion-nucleon transition matrix in the $P_{33}$ partial wave is
\widetext
\begin{mathletters}
\begin{eqnarray}\label{3.5a}
t(z_\Delta,k_\Delta)
&=&Qh_1(i)P_\Delta  \nonumber \\
& \times  &
\frac{P_\Delta}
{\displaystyle
z_\Delta-m_\Delta^0-\frac{k_\Delta^2}{2m_\Delta^0}-P_\Delta h_1(i)Q
\frac Q{z_\Delta-Qh_0(i)Q-Qh_0(\pi)Q}Qh_1(i)P_\Delta}P_\Delta h_1(i)Q\ ,
\nonumber \\
\\
t(z_\Delta,k_\Delta)&=&
|f\rangle \frac1{z_\Delta-M_\Delta(z_\Delta,k_\Delta)-\frac{k_\Delta^2}
{2m_\Delta^0}+\frac i2\Gamma_\Delta(z_\Delta,k_\Delta)}\langle f|
\label{3.5b}
\end{eqnarray}
\end{mathletters}
\narrowtext\noindent
with $Qh_1(i)P_\Delta=|f\rangle$, $P_\Delta h_1(i)Q=\langle f|$.
Pion-nucleon relative and c.m. momenta, i.e., $\bbox{\pi}$ and
${\bf k}_\Delta$, are introduced by
\begin{mathletters}
\begin{eqnarray}
\bbox{\pi}&=&\frac{\omega_\pi(k_\pi){\bf k}_N-m_N{\bf k}_\pi}
{m_N+\omega_\pi(k_\pi)}\ ,\\
{\bf k}_\Delta&=&{\bf k}_N+{\bf k}_\pi\ .
\end{eqnarray}
Since the pion is treated relativistically and the nucleon
nonrelativistically,
the reduction of operators from a many-baryon to a single-baryon form and,
conversely, the embedding of a single-baryon operator in many-baryon
systems can often
be done only approximately. We use the approximation
\begin{eqnarray}
&&\hspace{-0.7cm} Qh_0(i)Q+Qh_0(\pi)Q \nonumber \\
&\hspace{-0.7cm} =&\nonumber
m_N+\frac{k_N^2}{2m_N}+\sqrt{m_\pi^2+k_\pi^2}\\
\nonumber
&\hspace{-0.7cm} \approx&
m_N+\frac{\pi^2}{2m_N}+\sqrt{m_\pi^2+\pi^2}+
\frac{k_\Delta^2}{2(m_N+\sqrt{m_\pi^2+\pi^2})}\\
&\hspace{-0.7cm} =&
Qh^{\pi N}_{0\mbox{\tiny\
rel}}(i)Q+\frac{k_\Delta^2}{2(m_N+\sqrt{m_\pi^2+\pi^2})}\ ,
\end{eqnarray}
\end{mathletters}
which avoids angles between the three-momenta $\bbox{\pi}$ and ${\bf k}_\Delta$
in
the kinetic energy operator; it is employed
in the step from Eq. (\ref{3.5a}) to Eq.
(\ref{3.5b}), and it is believed to be quite accurate.
The kinetic energy operator of relative pion-nucleon motion
$Qh^{\pi N}_{0\mbox{\tiny\ rel}}(i)Q$ is introduced.
Thus, the effective mass $M_\Delta(z_\Delta,k_\Delta)$ and
the effective width $\Gamma_\Delta(z_\Delta,k_\Delta)$ of the
$\Delta$-isobar with momentum ${\bf k}_\Delta$
yield that pionic correction $P_\Delta\delta H_0(z)P_\Delta$, once the
propagation of two additional nucleons in $P_\Delta\delta H_0(z)P_\Delta$
is taken into account according to Eq.~(\ref{3.2a}). In a
three-baryon system with single $\Delta$-isobar excitation
\widetext
\begin{mathletters}
\begin{eqnarray}
\label{3.7a}
&&P_\Delta[H_0+\delta H_0(z)]P_\Delta|p_\Delta q_\Delta\nu_\Delta
\rangle_1=\left[2m_N+\frac{p_\Delta^2}{m_N}+\frac{q_\Delta^2}{4m_N}
+M_\Delta\left(\mbox{$z-2m_N-\frac{p_\Delta^2}{m_N}
-\frac{q_\Delta^2}{4m_N}$},
q_\Delta
\right)
\right.
\nonumber\\
&&\left.
+\frac{q_\Delta^2}{2m_\Delta^0}
-\frac i2\Gamma\left(\mbox{$z-2m_N-\frac{p_\Delta^2}{m_N}-
\frac{q_\Delta^2}{4m_N}$},q_\Delta\right)\right]|p_\Delta q_\Delta\nu_\Delta
\rangle_1\ .
\end{eqnarray}
\narrowtext\noindent
This paper carries out a trinucleon bound-state calculation only, thus,
the required three-baryon available energy $z$ is always smaller than
$3m_N$. As a consequence, the single-baryon available energy in the
effective mass and width of the $\Delta$-isobar according to
Eq. (\ref{3.7a}) is with $z-2m_N-p_\Delta^2/m_N-q_\Delta^2/
4m_N<m_N<m_N+m_\pi$. The width $\Gamma_\Delta(z_\Delta,k_\Delta)$
therefore does not contribute to the effective three-baryon resolvent
(\ref{2.5}), which in the Hilbert sector  ${\cal H}_\Delta$ takes the form
\widetext
\begin{eqnarray}\label{3.7b}
&&{}_1\langle p_\Delta'q_\Delta'\nu_\Delta'|G_0^P(z)
|p_\Delta q_\Delta\nu_\Delta\rangle_1=
\frac{\delta(p_\Delta'-p_\Delta)}{p_\Delta^2}\frac{\delta(q_\Delta'-q_\Delta)}
{q_\Delta^2}\delta_{\nu_\Delta'\nu_\Delta}
\nonumber\\
&& \times
\left[\mbox{$
z-2m_N-\frac{p_\Delta^2}{m_N}-\frac{q_\Delta^2}2\left(\frac1{2m_N}+
\frac1{m_\Delta^0}\right)-M_\Delta\left(z-2m_N-\frac{p_\Delta^2}{m_N}-
\frac{q_\Delta^2}{4m_N},q_\Delta\right)$}\right]^{-1}
\end{eqnarray}
\end{mathletters}
\narrowtext\noindent
for the considered available energies $z$ and for the basis states
$|p_\Delta q_\Delta\nu_\Delta\rangle_1$ of Eq.~(3.4).

The trinucleon bound-state calculation requires the operators in the
Hilbert sector ${\cal H}_\Delta$ also in the basis $|p_Nq_N\nu_N\rangle_1$
of Eq.~(3.6).
Compared with Eq.~(\ref{3.7b}) the three-baryon resolvent is more complicated,
since nondiagonal, in this basis. In order to
simplify calculations, we opted for an approximation which also makes
${}_1\langle p_N'q_N'\nu_N'|G_0^P(z)|p_Nq_N\nu\rangle_1$ diagonal in the
momenta and channels
of the basis states $|p_Nq_N\nu_N\rangle_1$. The approximation is best seen
in the quantity
\widetext
\begin{eqnarray}
\label{3.10}
& & \frac1{z-QH_0Q}QH_1P_\Delta|p_Nq_N\nu_N\rangle_1  \nonumber \\
&  =  &
\frac Q{[z-Qh_0(1)Q-Qh_0(2)Q]-Qh_0(3)Q-Qh_0(\pi)Q}Qh_1(3)P_\Delta|p_Nq_N
\nu_N\rangle_1
 \nonumber \\
&=&\nonumber
\frac1{z-2m_N-\frac{k_{N1}^2}{2m_N}-\frac{k_{N2}^2}{2m_N}-\frac{k_\Delta^2}
{2m_\Delta^0}-\left[Qh_0(3)Q-Qh_0(\pi)Q-\frac{k_\Delta^2}{2m_\Delta^0}\right]}
Qh_1(3)P_\Delta|p_Nq_N\nu_N\rangle_1\\
&\approx&
\frac1{z-2m_N-\frac{k_{N1}^2}{2m_N}-\frac{k_{N2}^2}{2m_N}-\frac{k_\Delta^2}
{2m_\Delta^0}-Qh^{\pi N}_{0\mbox{\tiny\ rel}}(3)Q}
Qh_1(3)P_\Delta|p_Nq_N\nu_N\rangle_1\ .
\end{eqnarray}
\narrowtext\noindent
The $\Delta$-isobar carries the baryon label 3.
The baryon kinetic
energy operator $k_{N1}^2/2m_N+k_{N2}^2/2m_N+k_\Delta^2/2m_\Delta^0$
without rest masses
is rewritten in terms of the Jacobi momenta $({\bf p}_N{\bf q}_N)$
of Eq.~(3.5),
whereas $\left[Qh_0(3)Q-Qh_0(\pi)Q-k_\Delta^2/2m^0_\Delta\right]$ is
approximated in the last step of Eq.~(3.10)
by the pion-nucleon relative kinetic energy
$Qh^{\pi N}_{0\mbox{\tiny\ rel}}(3)Q$ including rest
masses. According to Eq.~(3.8c)
$Qh_0(3)Q+Qh_0(\pi)Q$
has a c.m. contribution $k_\Delta^2/2\left(m_N+\sqrt{m_\pi^2+\pi^2}\right)$
with $k_\Delta^2$ {\it not} being a function of the magnitudes $p_N$ and $q_N$
of the Jacobi momenta only, but also of the angle between them;
the c.m. dependence couples partial
waves $\nu_N$ in a nontrivial way. Once that c.m. contribution
is accounted for by $k_\Delta^2/2m_\Delta^0$ with sufficient
accuracy, the basis states $|p_Nq_N\nu_N\rangle_1$ become true eigenstates of
the single-baryon part $H_0+\delta H_0(z)$ in the effective hamiltonian, i.e.,
\begin{mathletters}
\widetext
\begin{eqnarray}
&&P_\Delta\left[H_0+\delta H_0(z)\right]P_\Delta|p_Nq_N\nu_N
\rangle_1
=\left[2m_N+\frac{p_N^2}2\left(\frac1{m_N}+\frac1{m_\Delta^0}
\right)+\frac{q_N^2}2\left(\frac1{m_N+m_\Delta^0}+\frac1{m_N}\right)\right.
\nonumber\\
&&\nonumber+\left.
M_\Delta\left(z-2m_N-\frac{p_N^2}2\left(\frac1{m_N}+\frac1{m_\Delta^0}\right)
-\frac{q_N^2}2\left(\frac1{m_N+m_\Delta^0}+\frac1{m_N}\right),0\right)
\right.\\
&&\left.
-\frac i2\Gamma_\Delta\left(z-2m_N-\frac{p_N^2}2
\left(\frac1{m_N}+\frac1{m_\Delta^0}\right)-\frac{q_N^2}2\left(\frac1{m_N+
m_\Delta^0}+\frac1{m_N}\right),0\right)\right]|p_Nq_N\nu_N\rangle_1\ .
\nonumber \\
\end{eqnarray}
The approximation (3.10) employs the effective mass
$M_\Delta(z_\Delta,k_\Delta)$ and the effective width $\Gamma_\Delta(z_\Delta,
k_\Delta)$ of the $\Delta$-isobar at the momentum $k_\Delta=0$, though the
pion-nucleon system is not at rest in a three-baryon system; the reason is that
the approximation (3.10) works with the pion-nucleon relative kinetic
energy operator $Qh^{\pi N}_{0\mbox{\tiny\ rel}} Q$ and pushes the dependence
on the moving pion-nucleon c.m. into an appropriate available
energy $z_\Delta$.
The approximation (3.10) also assumes baryon 3 to be the $\Delta$-isobar;
however, the result (3.11a) is symmetric in baryons 2 and 3 and therefore
applies to the basis state $|p_Nq_N\nu_N\rangle_1$ antisymmetrized with
respect to baryons 2 and 3. Using the result (3.11a) the
effective three-baryon resolvent (\ref{2.5}) becomes diagonal also in
the basis states $|p_Nq_N\nu_N\rangle_1$ of the Hilbert
sector ${\cal H}_\Delta$
and takes the form
\begin{eqnarray}
& & {}_1\langle p_N'q_N'\nu_N'|G_0^P(z)|p_Nq_N\nu_N\rangle_1 =
\frac{\delta(p_N'-p_N)}{p_N^2}
\frac{\delta(q_N'-q_N)}{q_N^2}\delta_{\nu'_N\nu_N}\nonumber\\
&  \times &
\left[\mbox{$
z-2m_N- \frac{p_N^2}{2}
\left(\frac{1}{m_N} + \frac{1}{m_\Delta^0}\right)
-\frac{q_N^2}{2}
\left(\frac{1}{m_N+m_\Delta^0} + \frac{1}{m_N}\right)
$}\right. \nonumber\\
&- & \left.
\mbox{$
 M_\Delta
\left(z-2m_N-\frac{p_N^2}{2}
\left(\frac{1}{m_N}+\frac{1}{m_\Delta^0}\right)
-\frac{q_N^2}{2}
\left(\frac{1}{m_N+m_\Delta^0} + \frac{1}{m_N}\right),0\right)
$}
\right]^{-1}
\end{eqnarray}
\end{mathletters}
for the available energy $z$ considered in the trinucleon bound state.

Due to
its energy dependence, the $\Delta$-isobar self-energy correction
$P_\Delta\delta H_0(z)P_\Delta$ is different in a two-baryon system: It notices
the absence of the third baryon. When described by the basis states
$|p_N\nu_N\rangle$ for particles 2 and 3 in the basis of Eq.~(3.6),
it becomes
\begin{eqnarray}\label{3.12}
P_\Delta
\left[H_0+\delta H_0^{[2]}(z_{N\Delta})\right]
|p_N\nu_N\rangle
& = &
\left[m_N+\frac{p_N^2}2
\left(\frac1{m_N}+\frac1{m_\Delta^0} \right)
\right. \nonumber\\
& + &
 M_\Delta
\left(z_{N\Delta}-m_N-\frac{p_N^2}2
\left(\frac1{m_N}+\frac1{m_\Delta^0}\right)
,0\right)
\nonumber \\
& - & \left.
\frac i2\Gamma_\Delta
\left(z_{N\Delta}-m_N-
\frac{p_N^2}2
\left(\frac1{m_N}+\frac1{m_\Delta^0}\right)
,0\right)
\right]|p_N\nu_N\rangle
\end{eqnarray}
\narrowtext\noindent
The approximation of Eq.~(\ref{3.10}) on the pion-nucleon c.m. kinetic
energy is used accordingly. The superscript $[2]$ indicates that the
self-energy correction $P_\Delta\delta H_0^{[2]}(z_{N\Delta})P_\Delta$ refers
to a two-baryon c.m. system; it will be
needed for the actual realization of the
employed force model in Subsect.~D.

The validity of the approximation (3.10), important for the three-baryon
resolvent (3.11b) and for the self-energy correction (3.12) in the
two-baryon system, is proven as follows: The approximating step in
Eq.~(3.10) is not carried out,
$[Qh_0(3)Q+Qh_0(\pi)Q-{\bf k}_\Delta^2/(2m_\Delta^0)]$ is used full, the
dependence of ${\bf k}_\Delta$ on the angle between ${\bf p}_N$ and ${\bf q}_N$
is
kept; however, that angle is assumed to be fixed at values $0$ or
$\pi/2$ or $\pi$, respectively, thus, ${\bf k}_\Delta$ again remains
effectively dependent on the magnitudes of the Jacobi momenta $p_N$ and $q_N$
only and does not yield any channel coupling. Using approximation (3.10) and
the three different approximations indicated in this paragraph in
trinucleon calculations, the trinucleon binding energy varies by less than
1 keV, i.e., within the numerical accuracy; the approximation (3.10) is
henceforth considered quite satisfactory.
\subsection{Retarded One-Pion Exchange $P_\Delta\delta H_1(z)P_\Delta$}
The retarded one-pion exchange $P_\Delta\delta H_1(z)P_\Delta$ of the
effective baryonic hamiltonian in Eq.~(\ref{2.3a}) is illustrated by
process (b) of Fig.~4. It is defined in Eq.~(\ref{3.2b}) for the
three-baryon system. Refs~\cite{4,5} test the employed force model
in the two-nucleon system above pion threshold. Thus, the retarded one-pion
exchange $P_\Delta\delta H_1(z)P_\Delta$ is also needed in the
two-baryon c.m. system. It is different there, since it notices the absence
of the third noninteracting nucleon. It is notationally differentiated
as $P_\Delta\delta H_1^{[2]}(z_{N\Delta})P_\Delta$ by the superscript $[2]$.
It takes the form
\widetext
\begin{mathletters}
\begin{eqnarray}
P_\Delta\delta H_1^{[2]}(z_{N\Delta})P_\Delta
&=&v^{N\Delta\to\Delta N}(z_{N\Delta})\ ,\\
v^{N\Delta\to\Delta N}(z_{N\Delta})&=&\sum_{{j,k=2,3}\atop{j\neq k}}
P_\Delta h_1(k)Q
\nonumber \\
& \times &
\frac Q{ z_{N\Delta}-\left[Qh_0(2)Q+Qh_0(3)Q+
Qh_0(\pi)Q-\frac{\left({\bf k}_{N2}+{\bf k}_{N3}+{\bf
k}_\pi\right)^2}{2m_N+\sqrt{
m_\pi^2+k_\pi^2}}\right]}Qh_1(j)P_\Delta\ .
\nonumber \\
\end{eqnarray}
\end{mathletters}
\narrowtext\noindent
It can be calculated in the basis $|p_N\nu_N\rangle$ used in Eq.~(\ref{3.12})
for the corresponding $\Delta$-isobar self-energy correction
$P_\Delta\delta H_0^{[2]}(z_{N\Delta})
P_\Delta$ in the two-baryon system; the explicit form of its matrix elements
is given in Ref.~\cite{14}.
When embedding the retarded one-pion exchange into a three-baryon system,
the approximation $m_N+\sqrt{m_\pi+k_\pi^2}\approx m_\Delta^0$ is used as in
Eq.~(3.10) for
the total mass of the interacting
pion-nucleon system. Its relation to the same process in the two-nucleon
system can then be given, i.e.,
\widetext
\begin{eqnarray}
& &
{}_1\langle p_N'q_N'\nu_N'|P_\Delta\delta H_1(z)P_\Delta|p_Nq_N\nu_N\rangle_1
=\frac{\delta(q_N'-q_N)}{q_N^2}
\nonumber \\
& \times &
\langle p_N'\nu_N'|\mbox{$v^{N\Delta\to\Delta N}\left(z-m_N-
\frac{q_N^2}2\left(\frac1{m_N}+\frac1{m_N+m_\Delta}\right)\right)$}
|p_N\nu_N\rangle\ .
\end{eqnarray}
\narrowtext\noindent
The three-baryon basis $|p_Nq_N\nu_N\rangle_1$ of Eq.~(3.5)
is the appropriate
one for a nucleon-$\Delta$ interaction in a three-baryon system. We note
that the energy dependence in the retarded one-pion exchange $P_\Delta\delta
H_1(z)P_\Delta$ has a precise meaning and changes the retarded interaction in
a controlled way depending on the many-baryon system into which it is embedded.
\subsection{Parametrization of the Interaction Hamiltonian (2.1)}\label{sec3.3}
The interaction hamiltonian
is the force model of Fig.~\ref{fig2} with active pion and $\Delta$-isobar
degrees of freedom. Those degrees of freedom only become active in
isospin-triplet partial waves; in isospin-singlet partial waves the
interaction is purely nucleonic and represented solely by process (a)
of Fig.~\ref{fig2}.
In this paper it is assumed that the interaction hamiltonian $H_1$
vanishes in the pionic sector, i.e., $QH_1Q=0$. In general, that is a
physically severe assumption, employed already in Ref.~\cite{4}:
E.g., in the presence of a pion two nucleons cannot be bound; thus, all
pion-deuteron processes are not described internally consistent under such
an assumption.

Two distinct parametrizations, labelled by $H(1)$ and $S(1)$ in the tables
and the result section, are chosen for the baryonic interaction,
which differ by their forms of the pion- and rho-exchange transition
potential $P_\Delta H_1P_N$: The parametrization with the transition
potential of Ref.~\cite{10} used there for the force model is labelled
$H$ here, since it is based on rather {\it hard}
form factors,
whereas that with the transition potential of Refs.~\cite{4,5} based on
rather {\it soft} form factors is labelled by $S$.

The instantaneous
nucleon-$\Delta$ potential $P_\Delta H_1P_\Delta$ is that
of Refs.~\cite{4,17};
its exchange part illustrated in Fig.~\ref{fig2}(c)
is based on pion and rho exchange; only half of the full pion exchange
is kept in $P_\Delta H_1P_\Delta$, since the other half, denoted by $\frac12
\pi_R$ in Table~\ref{tab1}, is generated explicitly
by the force model as $P_\Delta\delta H_1(z)P_\Delta$ in a retarded fashion
according to Eq.~(3.14); the subscript $R$ in the notation $\frac12\pi_R$
indicates its retardation. The unretarded half of the pion exchange, kept in
$P_\Delta H_1P_\Delta$, is identified with the on-shell form
$v^{N\Delta\to\Delta N}(z_{N\Delta\mbox{\tiny on}})$ of Eq.~(3.13a) as in
Refs.~\cite{14,4}; it is denoted
by $\frac12\pi_S$ in Table~\ref{tab1}, the subscript $S$ in the notation
$\frac12\pi_S$ indicates that it is unretarded, but based on the soft
form factors of the retarded pion-exchange $P_\Delta\delta H_1(z)P_\Delta$.

In contrast to Refs.~\cite{4,5} the nucleonic
part of the interaction is chosen as
\widetext
\begin{equation}
P_NH_1P_N=V_{NN}\label{3.11a}-P_NH_1P_\Delta\frac{P_\Delta}
{2m_N-P_\Delta\left[H_0+
\delta H^{[2]}_0(2m_N)+H_1
+\delta H_1^{[2]}(2m_N)\right]P_\Delta}P_\Delta H_1P_N
\end{equation}
\narrowtext\noindent
The choice (\ref{3.11a}) yields exact phase
equivalence at zero kinetic energy and
approximate phase equivalence at low kinetic
energies between the full force model
and a realistic, but purely nucleonic reference potential $V_{NN}$.
The Paris Potential \cite{18} is chosen as reference potential $V_{NN}$.
That reference potential is employed in all isospin-singlet partial waves.
The choice (\ref{3.11a}) is a conceptual improvement compared to
Ref.~\cite{4}. The improved phase equivalence is documented in
Ref.~\cite{19} which also demonstrates that that improvement is
quantitatively irrelevant for observables of the two-nucleon system above
pion threshold.
\subsection{Solution of Trinucleon Equations}\label{sec3.4}
The Faddeev equations (\ref{3.1}) are solved in momentum space
using the technical apparatus
of Refs.~\cite{10} and \cite{13}. The two-baryon interaction is assumed to act
in all partial waves up to total pair angular momentum $I=2$. 18 purely
nucleonic Faddeev amplitudes $P_N|\psi_i\rangle $ arise in the partial-wave
decomposition defined by the Jacobi coordinates and discrete quantum numbers
of
Fig.~\ref{fig5}; in addition, 14 Faddeev amplitudes
$P_\Delta|\psi_i\rangle $ with a single $\Delta$-isobar in the pair and one
with the $\Delta$-isobar as
spectator are taken into account as in Refs.~\cite{10,13}. The
employed discretization of the equations (\ref{3.1})
is the one of Ref.~\cite{13}. The triton
binding energy $E_T$, and the baryonic
components $P|\Psi_B\rangle $ of the wave
function according to Eq.~(\ref{2.9b}), are obtained from such a calculation
for the force model defined in Subsect.~D. The
pionic component $Q|\Psi_B\rangle $ of the wave function
can in principle be gotten from the
baryonic ones $P|\Psi_B\rangle$ by Eq.~(\ref{2.3b});
however, this paper only computes
its weight $\langle \Psi_B|Q|\Psi_B\rangle $ in the
trinucleon wave function according to
\widetext
\begin{mathletters}
\begin{eqnarray}
& & \langle\Psi_B|Q|\Psi_B\rangle  =
\langle \Psi_B|P_\Delta H_1Q\frac Q{(E_B-QH_0Q)^2}QH_1P_\Delta|
\Psi_B\rangle \nonumber \\
& & =  \left.
\langle \Psi_B|P_\Delta\left(-\frac\partial{\partial z}\right)
\left[\delta H_0(z)+\delta H_1(z)\right]P_\Delta|\Psi_B\rangle
\right|_{z=E_B}
\\
& &\langle\Psi_B|Q|\Psi_B\rangle  =
3 \sum_\nu\int p_\Delta^2dp_\Delta q_\Delta^2dq_\Delta
\langle \Psi_B|p_\Delta q_\Delta\nu_\Delta\rangle_1
\nonumber \\
& & \times
\left.
\left(-\frac\partial{\partial z_\Delta}
\right)M_\Delta\left(z_\Delta,q_\Delta\right)\
{}_1\langle p_\Delta q_\Delta\nu_\Delta|\Psi_B\rangle
\right|_{z_\Delta
=E_T+m_N-\frac{p_\Delta^2}{m_N}-\frac{q_\Delta^2}{4m_N}}
\nonumber \\
& & +
3\sum_{\nu'\nu}\int p_N'^2dp_N'p_N^2dp_Nq_N^2dq_N
\langle \Psi_B|p_N'q_N'\nu_N'\rangle _1
\nonumber \\
& & \times
\left.\left(-\frac\partial{\partial z_{N\Delta}}\right)
\langle p_N'\nu'_N|
v^{N\Delta\to\Delta N}(z_{N\Delta})|p_N\nu_N\rangle {}_1\langle p_Nq_N\nu_N
|\Psi_B\rangle
\right|_{z_{N\Delta}
=E_T+2m_N-\frac{q_N^2}2\left(\frac1{m_N}+\frac1{m_N+m_\Delta^0}
\right)}\ . \nonumber \\
\end{eqnarray}
\end{mathletters}
\narrowtext
\subsection{Comparison with Coupled-Channel Calculations}
The force model of this paper, defined in Fig.~\ref{fig2} and employed
in the calculation of the trinucleon bound state, is used in two
parametrizations as Subsect.~D describes. In the force model,
the bare $\Delta$-isobar
is dynamically coupled to pion-nucleon states and builds up, by that coupling,
the physical $P_{33}$ resonance in the nuclear medium. We say the
force model is based on a {\it dynamic} $\Delta$-isobar.
The calculation of that full force model is
compared with coupled-channel calculations of the trinucleon bound state
in which the $\Delta$-isobar does not couple to pion-nucleon states, keeps
a fixed mass $m_\Delta^0$ without any pionic correction, i.e.,
$P_\Delta\delta H_0(z)P_\Delta=0$,
and interacts with a nucleon only through unretarded
potentials, i.e., $P_\Delta\delta H_1(z)P_\Delta=0$.
In those coupled-channel calculations $QH_1P_\Delta=P_\Delta H_1Q=0$, thus,
the trinucleon bound state does not have any pionic components either, i.e.,
$Q|\Psi_B\rangle=0$. We say those coupled-channel variants of the full
force model are based on a {\it stable} $\Delta$-isobar.
The technical apparatus which this paper borrows from
Refs.~\cite{10,13} was originally designed for those coupled-channel
calculations.

The aim of this paper is to find the validity of coupled-channel approximations
to the full force model. This is the reason why various choices for the
stable $\Delta$-isobar mass $m_\Delta^0$ and for the instantaneous pion
exchange between the $\Delta$-isobar and a nucleon are tried out. The various
choices are listed in Table~\ref{tab1} and are described -- together with
the trinucleon results -- in Sect.~IV.

\section{Results} \label{sec4}

Calculations of the form which Refs.~\cite{4,5} and this paper report on have
three physics objectives in mind:
\begin{enumerate}
\item The full force model of Fig.~\ref{fig2}
should be tuned in the two-nucleon
system above pion threshold. In particular, that tuning process should fix the
strength and shape of the two-baryon transition potential
$P_\Delta H_1P_N$ from two-nucleon to nucleon-$\Delta$ states and of the
nucleon-$\Delta$ potential $P_\Delta H_1P_\Delta$, potentials on which one
lacks detailed information otherwise.
\item The calculation of the trinucleon bound state should determine the
amount of three-nucleon force arising from the explicit excitation of a
$\Delta$-isobar and
from the explicit production of a pion in the trinucleon bound state.
\item The conditions under which the simpler coupled-channel calculations
for the trinucleon bound state approximate the results of the full force
model with a dynamic $\Delta$-resonance are to be found.
\end{enumerate}
This paper follows that ambitious program, though it is unable to carry it
through to full satisfaction.
It uses two versions of the force model of Fig.~\ref{fig2} with
the approximation $QH_1Q=0$. That approximation is
an inconsequential one for the
trinucleon bound state, as the results will prove,
but a fatal one, if serious tuning of the force
model to all observables in the two-nucleon system above pion threshold
were attempted.

The version $S$ based on soft form factors in the transition potential
is tested for many observables of the two-nucleon system above pion threshold
and its successes and failures are well documented in
Refs.~\cite{4,5};
version $S$ is a moderatly realistic force model. In contrast,
the version $H$ based on hard form factors in the transition potential has
not been tested yet; its realistic nature above pion threshold
is doubtful; version
$H$ is used in this paper, since the original coupled-channel calculations
\cite{9,10} for the trinucleon bound state were based on its transition
potential. Thus, item (1) of the program list is not carried out with
any satisfaction. The two versions are labelled $H(1)$ and $S(1)$ in
Table~\ref{tab1} which summarizes their defining properties and in
Table~\ref{tab2} which collects their predictions for the triton binding
energy $E_T$, for the effective two-nucleon and three-nucleon contributions
of the $\Delta$-isobar and the pion to the binding, i.e., $\Delta E_2$ and
$\Delta E_3$ according to the technique of Ref.~\cite{10},
and for the nucleonic, $\Delta$-isobar and pionic probabilites
in the triton bound state, i.e., $P_{{\cal L}},\ P_\Delta$ and $P_\pi$.
The effect of the explicit $\Delta$-isobar
{\it and} pion degrees of freedom on the trinucleon bound state properties
are for the chosen versions of the full force model well isolated. Thus,
item (2) of the program list is carried out.

The results derived from the full force models $H(1)$ and $S(1)$ are
compared with those of corresponding standard coupled-channel
calculations without explicit pion degrees of freedom. In those
coupled-channel calculations the effective $\Delta$-mass $M_\Delta(z_\Delta,
k_\Delta)$ is taken to be constant and is equated to a stable mass
$m_\Delta^0$, its standard value being 1232 MeV, the resonance position
in $P_{33}$ pion-nucleon scattering. Furthermore, the pionic nucleon-$\Delta$
exchange potential becomes instantaneous. In the standard coupled-channel
calculation of Ref.~\cite{17} that pionic contribution is used in a local form
with hard form factors, denoted by $\pi_H$ in Table~\ref{tab1}.
The two standard coupled-channel force models
are labelled $H(4)$ and $S(4)$, $H(4)$ is identical with the force model
A3 of Ref.~\cite{17}.
Table \ref{tab1} summarizes the defining properties of the full force model
in the parametrizations $H(1)$ and $S(1)$ and three coupled-channel variants
of the full force model, i.e., $H(2),\ H(3),\ H(4)$ and
$S(2),\ S(3),\ S(4)$; the variants (2) and (3) interpolate between the full
force models (1) and the standard coupled-channel models (4):
Variant (2) works with a mass of 1290 MeV for the stable $\Delta$-isobar
which is larger than the resonance value 1232 MeV; the effective $\Delta$-mass
$M_\Delta(z_\Delta,k_\Delta)$ is documented in Ref.~\cite{2}, it is not
shown again in this paper; in bound-state problems the available energy
$z_\Delta$ of the effective $\Delta$-mass is smaller than $m_N$ according
to Sect. IIIB and then the effective $\Delta$-mass becomes larger than its
resonance value 1232 MeV and approaches the bare mass of 1315 MeV; thus, the
value of 1290 MeV chosen as stable $\Delta$-mass in variant (2) should
approximate the effective $\Delta$-mass $M_\Delta(z_\Delta,k_\Delta)$
rather well; variant (2) also works with the instantaneous limit $\pi_S$
for the pion-exchange nucleon-$\Delta$ potential; variant (2) should
reproduce the results of the full force models $H(1)$ and $S(1)$ best. Variant
(3) works with a stable $\Delta$-isobar mass of 1232 MeV, the resonance
value, but preserves $\pi_S$ for the pion-exchange nucleon-$\Delta$
potential. Table II summarizes the respective trinucleon results.
It is also worth noticing  at this point that in models (1), based on
a dynamic $\Delta$ isobar, from the two contributions to the
pionic probability in Eq. (3.16), the self-energy contribution $\delta H_0$
dominates over the retardation term $\delta H_1$, which is found to be
one order of magnitude lower.

The main result of this paper -- in answer to item (3) of the program
list -- is: The trinucleon properties
derived from the full force model, defined
in Fig.~\ref{fig2} and parametrized in Sect.~\ref{sec3}.D as $H(1)$ and $S(1)$,
are well approximated
by those of the corresponding standard coupled-channel models
$H(4)$ and $S(4)$ with a stable $\Delta$-isobar.
The standard coupled-channel models account
for
all corrections of trinucleon properties due to the explicit $\Delta$-isobar
and pion degrees of freedom within $90\%$.
The quality of the approximation can be
read of from
Table~\ref{tab2} where also the results of the Paris potential, the
purely nucleonic reference potential for all considered
force models, are listed. The quality of the approximation even improves
when the parametrization of the coupled-channel model is better tuned to
the full force model as for example in variant (2). On the other hand the fact
that the standard coupled-channel models $H(4)$ and $S(4)$ approximate some
trinucleon properties of their corresponding full force models even more
successfully than their seemingly better tuned variants (3) appears to be
accidental.

As expected, the force model $H(1)$ based on hard form factors in the
transition potential $P_\Delta H_1P_N$ to nucleon-$\Delta$ states yields
larger probabilities $P_\Delta$ and $P_\pi$ for the $\Delta$-isobar and the
pion in the trinucleon bound state than the force model $S(1)$ does with
its softer form factors. In both cases, however, the probability $P_\pi$
of the pionic components in the wave function is extremely small; thus, the
simplifying approximation $QH_1Q=0$ which neglects all interactions in the
pionic sector ${\cal H}_\pi$ of the Hilbert space is well justified.

\section{Conclusions}\label{sec5}

For the first time, this paper carries out the conceptual idea underlying
the previous coupled-channel calculations of Refs.~\cite{9}, \cite{10} and
\cite{17}:

A contribution to the three-nucleon force arises from the mechanism for pion
production and pion absorption; that mechanism is seen in the two-nucleon
system above threshold. This paper makes the step from two-nucleon reactions
without and with a pion to trinucleon properties and isolates effects related
to the explicit $\Delta$-isobar and pion degrees of freedom.

Furthermore, this paper justifies the general use of coupled-channel
calculations with stable $\Delta$-isobars and indicates ways for improving
their simulations of the full force model.
\acknowledgments
M.T.P. thanks the Theory Group at CEBAF for the
kind hospitality granted during her stay there.
This work was funded by the Deutsche Forschungsgemeinschaft (DFG) under
contract No. Sa 247/7-2 and Sa 247/7-3, by the Deutscher Akademischer
Austauschdienst (DAAD) under Contract No. 322-inida-dr, by the DOE under Grant
No. DE-FG05-88ER40435, and by JNICT under Contract No. PBIC/C/CEN/1094/92. The
calculations for
this paper were performed at Regionales Rechenzentrum f\"ur Niedersachsen
(Hannover),
at Continuous Electron Beam Accelerator Facility (Newport News),
and at National Energy Research Supercomputer Center (Livermore).

\appendix
\section{Calculation of the Effective Singular Three-Baryon Force
Arising from Processes of the Type Fig.~4(e)}\label{appA}
The process of Fig.~\ref{fig4}(e) is redrawn in Fig.~\ref{figA1};
a characteristic
process of higher order in potentials is also shown there. Characteristic
for both processes is that they are unlinked in a particular way: The two
nucleons unconnected with the $\Delta$-isobar can interact up to
infinite order of the two-nucleon potential; the nucleon produced by the
decay of the $\Delta$-isobar interacts with the simultaneously created
pion also up to infinite order in the pion-nucleon potential, however, it does
not interact with either of the other two nucleons. The processes of
Fig.~\ref{figA1} depend on the coordinates of all three baryons. They therefore
yield irreducible contributions to the effective three-baryon force. However,
due to that particular disconnectedness those contributions are singular
in the same way as a two-baryon interaction is singular in a three-baryon
Hilbert space. The disconnectedness problem, which the process of
Fig.~\ref{fig4}(e)
yields resembles the one encountered in two-nucleon scattering within
the framework of $\pi NN$ dynamics by Ref.~\cite{Stingl}.
This paper chooses $QH_1Q=0$, thus, that singular
three-baryon force does not arise in the actual calculation. Nevertheless,
its functional form is given in this appendix for conceptual completeness.

The interaction $QH_1Q$ in the Hilbert sector ${\cal H}_\pi$ with a pion is
decomposed as follows, i.e.,
\begin{equation}
QH_1Q=\sum_i v_i^{NN}+\sum_i v_i^{\pi N}\ .
\end{equation}
In the two-nucleon potential $v_i^{NN}$ the subscript $i$ denotes the
spectating nucleon, in the pion-nucleon potential $v_i^{\pi N}$ the subscript
$i$ denotes the nucleon interacting with the pion. All potentials in the
interaction $QH_1Q$ are instantaneous. Processes up to infinite order in
the potentials contribute; the potentials are resummed into transition
matrices, i.e.,
\widetext

\begin{mathletters}
\begin{eqnarray}
& &
t_1^{NN}\left(z-h_0(1)-h_0(\pi)-\frac{({\bf k}_{N2}+{\bf
k}_{N3})^2}{4m_N}\right)
\nonumber \\
& & \quad
= v_1^{NN}\left[1+\frac1{\left[z-h_0(1)-h_0(\pi)-\frac
{({\bf k}_{N2}+{\bf k}_{N3})^2}{4m_N}\right]-\left(h_0(2)+h_0(3)-\frac{
({\bf k}_{N2}+{\bf k}_{N3})^2}{4m_N}\right)}\right. \nonumber\\
& & \quad
\left.\phantom{\frac1{\frac{({\bf k}_{N2}+{\bf k}_{N3})^2}{4m_N}}}
\times  t_1^{NN}\left(
z-h_0(1)-h_0(\pi)-\frac{({\bf k}_{N2}+{\bf k}_{N3})^2}{4m_N}\right)\right]\ ,\\
& & t_1^{\pi N}\left(z-h_0(2)-h_0(3)\right)
\nonumber \\
& & \quad =
v_1^{\pi N}\left[
1+\frac1{
\left[z-h_0(2)-h_0(3)\right]-
h_0(1)-h_0(\pi)}\
t_1^{\pi N}\left(z-h_0(2)-h_0(3)\right)
\right]\ .
\end{eqnarray}
\end{mathletters}
\narrowtext

The two-nucleon transition matrix $t_1^{NN}$ is defined in the c.m. system
of nucleons 2 and 3; in Eq.~(A2a) the pion-nucleon kinetic energy will be
approximately split in the form $h_0(1)+h_0(\pi)=h_{0\
\mbox{\tiny rel}}^{\pi N}(1)+k_\Delta^2/2m_\Delta^0$ as in the context
of Eq.~(3.10). In contrast the pion-nucleon transition matrix
$t_1^{\pi N}$ is defined for a moving pion-nucleon system as for the
transition matrix in Eq.~(3.7); however different symbols are used for the
pion-nucleon transition matrices of Eqs.~(A2b) and (3.7), since their
dynamic content is different.

According to Eq.~(2.9a) only the part $W_1$ of the considered
three-baryon force is needed for determining the Faddeev amplitude
$P|\psi_1\rangle$; that part is calculated in the chosen basis of the
Hilbert sector ${\cal H}_\Delta$. However, the matrix elements
${}_1\langle p_N'q_N'\nu_N'|W_1(z)|p_Nq_N\nu_N\rangle_1$ and
the nondiagonal matrix elements ${}_1\langle p_\Delta q_\Delta\nu_\Delta|
W_1(z)|p_Nq_N\nu_N\rangle_1$ are identically zero, only the matrix elements
for the basis states $|p_\Delta q_\Delta\nu_\Delta\rangle_1$ are
nonzero and have to be computed. The operator $W_1$ of the considered
three-baryon force is build up by simpler quantities, i.e., the
two-nucleon and pion-nucleon transition matrizes $t_1^{NN}$ and
$t_1^{\pi N}$. Those transition matrices act in the Hilbert sector
${\cal H}_\pi$, they do not act on the three-baryon basis states
$|p_\Delta q_\Delta\nu_\Delta\rangle_1$, but these {\it outside}
basis states simplify those transition matrices
{\it inside} $W_1$  by allowing the replacement of some operators through
their corresponding eigenvalues, e.g.,

\widetext
\begin{mathletters}
\begin{eqnarray}\label{A3a}
& & t_1^{NN}\left(z-h_0(1)-h_0(\pi)
-\frac{({\bf k}_{N2}+{\bf k}_{N3})^2}{4m_N}\right)
 \ldots\ |p_\Delta q_\Delta\nu_\Delta\rangle_1
\nonumber \\
& & = t^{NN}
\left(z-h^{\pi N}_{0\mbox{\tiny\ rel}}(1)
-\frac{q_\Delta^2}2\left(\frac1{2m_N}+\frac1{m_\Delta^0}\right)\right)
\ldots |p_\Delta q_\Delta\nu_\Delta\rangle_1\ ,
\end{eqnarray}
\begin{equation}
t_1^{\pi N}\left(z-h_0(2)-h_0(3)
\right) \ldots|p_\Delta q_\Delta\nu_\Delta\rangle_1 \label{A3b}
=t^{\pi N}\left(z-2m_N-
\frac{p_\Delta^2}{m_N}-\frac{q_\Delta^2}{4m_N},\
q_\Delta\right) \ldots|p_\Delta q_\Delta\nu_\Delta
\rangle_1\ .
\end{equation}
\end{mathletters}
\narrowtext
\noindent
The dots in both equations indicate that the transition matrices do not act
directly on the basis states $|p_\Delta q_\Delta\nu_\Delta\rangle_1$.
Both transition matrices remain operators with respect
to the relative pion-nucleon motion.

The pion-nucleon transition matrix $t^{\pi N}$ of
Eq.~(\ref{A3b}) sums up the nonresonant part of the pion-nucleon interaction
in the $P_{33}$ partial wave. The nonresonant part is weak. Thus, only
contributions of first order in the pion-nucleon transition matrix $t^{\pi N}$
are
considered. The three arising contributions are shown in Fig.~\ref{figA2}.
They have the following analytic form
\widetext
\begin{eqnarray}\label{A4}
&&{}_1\langle p_\Delta'q_\Delta'\nu_\Delta'|W_1|p_\Delta q_\Delta\nu_\Delta
\rangle_1=
\nonumber\\
&&\frac{\delta(q_\Delta'-q_\Delta)}{q_\Delta^2}\mbox{\Large$\langle f|$}
\frac1{\left[z-2m_N-
\frac{p_\Delta'^2}{m_N}-\frac{q_\Delta^2}2\left(\frac1{2m_N}
+\frac1{m_\Delta^0}\right)\right]-h^{\pi N}_{0\mbox{\tiny\ rel}}(1)}
\nonumber\\
&&\times
\left\{
\langle p_\Delta'\nu_\Delta'|t^{NN}\left(z-\frac{q_\Delta^2}2\left(
\frac1{2m_N}+\frac1{m_\Delta^0}\right)-h^{\pi N}_{0\mbox{\tiny\ rel}}(1)
\right)|p_\Delta\nu_\Delta\rangle
\right.
\nonumber \\
& & \quad \times \left.
\frac1{\left[z-2m_N-\frac{p_\Delta^2}{m_N}
-\frac{q_\Delta^2}2\left(\frac1{2m_N}+\frac1{m_\Delta^0}\right)\right]
-h^{\pi N}_{0\mbox{\tiny\ rel}}(1)}\right.
\nonumber\\
\nonumber&&\quad \times
t^{\pi N}\left(\mbox{$z-2m_N-\frac{p_\Delta^2}{m_N}
-\frac{q_\Delta^2}2\left(\frac1{2m_N}+\frac1{m_\Delta^0}\right)$}
,q_\Delta\right)\\
&&\nonumber
+\sum_{\nu_\Delta''}\int p_\Delta''^2dp_\Delta''\langle p_\Delta'\nu_\Delta'|
t^{NN}\left(\mbox{$z
-\frac{q_\Delta^2}2\left(\frac1{2m_N}+\frac1{m_\Delta^0}\right)
-h^{\pi N}_{0\mbox{\tiny\ rel}}(1)$}\right)
|p_\Delta''\nu_\Delta''\rangle
\\&&\quad \times
\frac1{\left[z-2m_N-\frac{p_\Delta''^2}{m_N}
-\frac{q_\Delta^2}2\left(\frac1{2m_N}+\frac1{m_\Delta^0}\right)\right]
-h^{\pi N}_{0\mbox{\tiny\ rel}}(1)}
\nonumber \\
& & \quad \times
t^{\pi N}\left(\mbox{$z-2m_N-\frac{p_\Delta''^2}{m_N}
-\frac{q_\Delta^2}2\left(\frac1{2m_N}+\frac1{m_\Delta^0}\right)$}
,q_\Delta\right)
\nonumber \\\nonumber&&\quad \times\frac1{\left[z-2m_N-\frac{p_\Delta''^2}{m_N}
-\frac{q_\Delta^2}2\left(\frac1{2m_N}+\frac1{m_\Delta^0}\right)\right]
-h^{\pi N}_{0\mbox{\tiny\ rel}}(1)}
\nonumber \\
&&\quad\times
\langle p_\Delta''\nu_\Delta''|
t^{NN}\left(\mbox{$z
-\frac{q_\Delta^2}2\left(\frac1{2m_N}+\frac1{m_\Delta^0}\right)
-h^{\pi N}_{0\mbox{\tiny\ rel}}(1)$}\right)
|p_\Delta\nu_\Delta\rangle
\nonumber \\
& &+t^{\pi N}\left(\mbox{$z-2m_N-\frac{p_\Delta'^2}{m_N}
-\frac{q_\Delta^2}2\left(\frac1{2m_N}+\frac1{m_\Delta^0}\right)$}
,q_\Delta\right)
\nonumber\\
&&\quad\times
\frac1{\left[z-2m_N-\frac{p_\Delta'^2}{m_N}
-\frac{q_\Delta^2}2\left(\frac1{2m_N}+\frac1{m_\Delta^0}\right)\right]
-h^{\pi N}_{0\mbox{\tiny\ rel}}(1)}
\nonumber \\
& & \quad \times \left.
\langle p_\Delta'\nu_\Delta'|
t^{NN}\left(\mbox{$z
-\frac{q_\Delta^2}2\left(\frac1{2m_N}+\frac1{m_\Delta^0}\right)
-h^{\pi N}_{0\mbox{\tiny\ rel}}(1)$}\right)
|p_\Delta\nu_\Delta\rangle \right\}
\nonumber \\
&&\qquad\qquad\times
\frac1{\left[z-2m_N-\frac{p_\Delta^2}{m_N}
-\frac{q_\Delta^2}2\left(\frac1{2m_N}+\frac1{m_\Delta^0}\right)\right]
-h^{\pi N}_{0\mbox{\tiny\ rel}}(1)}\mbox{\Large$|f\rangle$}\ .
\end{eqnarray}
\narrowtext\noindent
The $\delta$-function $\delta(q_\Delta'-q_\Delta)/q_\Delta^2$ yields the
singular structure of the three-baryon force $W_i(z)$.

In calculations with $QH_1Q\neq0$ the effective
three-baryon force $W_i(z)$ arises and has the discussed singular part of
Eq.~(\ref{A4}). That singular part has to be combined with the two-baryon
interaction $v_i(z)$ of same singularity structure.



\newpage
\begin{figure}
\caption{Hilbert space for a many-nucleon system. Besides the
purely nucleonic sector ${\cal H}_N$ there is the sector ${\cal H}_\Delta$
in which one nucleon is turned into a $\Delta$-isobar and the sector
${\cal H}_\pi$ in which a single pion is added.}\label{fig1}
\end{figure}
\begin{figure}
\caption{Building blocks of the force model with $\Delta$-isobar and
pion degrees of freedom. The hermitian adjoint pieces corresponding
to the processes (b) and (e) are not shown. The $\Delta$-isobar
is a bare particle; process (e) yields the physical $P_{33}$
pion-nucleon resonance by iteration; process (f) stands for the
nonresonant pion-nucleon interactions; in general, it could also have
contributions in $P_{33}$; none of those possible $P_{33}$ background
contributions is indicated in the following Figs.~3 and 4.
The extended force model acts
in isospin-triplet partial waves only. In isospin-singlet partial
waves the force model is purely nucleonic and
reduces to process (a).}\label{fig2}
\end{figure}
\begin{figure}
\caption{Examples for contributions to the effective three-nucleon
force arising in a three-nucleon system
from the force model of Fig.~2.
The contributions are irreducible in the purely nucleonic Hilbert sector
${\cal H}_N$. Selected contributions up to fourth order in two-particle
potentials are shown. With respect to the pion-nucleon interaction,
possible $P_{33}$ background contributions are not considered in this
figure.}\label{fig3}
\end{figure}
\begin{figure}
\caption{Energy-dependent contributions to the effective
hamiltonian of Eq.~(2.3a) arising from projecting
out the pionic component from the trinucleon wave
function. All energy-dependent contributions act in the
baryonic Hilbert sector ${\cal H}_\Delta$ with a $\Delta$-isobar. In
the top row the only contribution of one-baryon
nature $P\delta H_0(z)P$ is shown. Row two (three) gives
characteristic examples of two-(three-)baryon
nature in $P\delta H_1(z)P$. With respect to the pion-nucleon
interaction, possible $P_{33}$ background contributions are not considered in
this figure. The contributions (c) -- (g) disappear, once interactions
in the Hilbert sector ${\cal H}_\pi$ are not taken into account, i.e.,
$QH_1Q=0$.}\label{fig4}
\end{figure}
\begin{figure}
\caption{Three-body Jacobi coordinates. The magnitude of the
corresponding momenta are $p_1$ and $q_1$. In the momentum-space
basis states $|p_1q_1\nu_1\rangle_1$ the antisymmetrized state of pair 2 and
3 and the spectator state are coupled with respect to their angular
momenta $I$ and $j$ and isospin $T$ and $t_1$ i.e.,
$|p_1q_1[(LS)I(ls_1)j]{\cal JJ}_z(Tt_1){\cal TT}_z\rangle_1$. The quantum
numbers $L(l)$ and $S(s_1)$ refer to the orbital angular momentum and spin
of the pair (spectator), ${\cal J}({\cal J}_z)$ and ${\cal T}({\cal T}_z)$
are total angular momentum (projection) and total isospin (projection)
of the three-body bound state.}\label{fig5}
\end{figure}

\begin{figure}
\caption{Effective three-baryon resolvent (2.5). Its form
in the Hilbert sector
${\cal H}_N$ is diagramatically shown
on the left side and its form in the Hilbert sector ${\cal H}_\Delta$ on
the right side.}\label{fig6}
\end{figure}
\begin{figure}
\caption{Characteristic
contribution to the pion-nucleon transition matrix in the
$P_{33}$ partial wave. The force model of Fig.~2 does not have
any additional background potential $QH_1Q$ in that partial wave. An
effective propagation of the $\Delta$-isobar can be read off from the
transition matrix and reoccurs -- together with the propagation of
two additional nucleons
-- in the effective three-baryon resolvent (2.5) of
Fig.~6.}\label{fig7}
\end{figure}
\begin{figure}
\caption{Characteristic
contributions to the effective three-nucleon force of
the type to be calculated in this appendix. The process of Fig.~4(e)
is redrawn; a characteristic process of higher order in the potentials, i.e.,
of third order in the two-nucleon potential and of second order in the
pion-nucleon potential, is also shown.}\label{figA1}
\end{figure}
\begin{figure}
\caption{The three
contributions of first order in the nonresonant pion-nucleon
transition matrix to the effective three-nucleon force, calculated in
this appendix. The shaded boxes denote nucleon-nucleon and pion-nucleon
transition matrices, they can be differentiated by their external legs.}
\label{figA2}
\end{figure}
\newpage
\mediumtext

\begin{table}
\caption{Employed Force Models with $\Delta$-Isobar and Pion Degrees of
Freedom}
\begin{tabular}{lccccr}
&Bare $\Delta$-Mass&Effective $\Delta$-Mass&$P_\Delta H_1 P_N$&
\multicolumn{2}{c} {$P_\Delta[H_1+\delta H_1(z)]P_\Delta$} \\
&&&& \multicolumn{2}{c} {$\pi$-Exchange, Fig.~\ref{fig2}(c)}\\
&$m_\Delta^0$ [MeV]&&&\\\tableline
$H$(1)&1315&$M_\Delta(z_\Delta,k_\Delta)$&$H$[10]&$\frac12\pi_R+\frac12\pi_S$ &
\cite{4}\\
$H$(2)&1290&$m_\Delta^0$&$H$[10]&$\pi_S$ & \cite{4}\\
$H$(3)&1232&$m_\Delta^0$&$H$[10]&$\pi_S$ & \cite{4}\\
$H$(4)&1232&$m_\Delta^0$&$H$[10]&$\pi_H$ & \cite{17}\\\hline
$S$(1)&1315&$M_\Delta(z_\Delta,k_\Delta)$&$S$\ [4]&$\frac12\pi_R+\frac12\pi_S$
&
\cite{4}\\
$S$(2)&1290&$m_\Delta^0$&$S$\ [4]&$\pi_S$ & \cite{4}\\
$S$(3)&1232&$m_\Delta^0$&$S$\ [4]&$\pi_S$ & \cite{4}\\
$S$(4)&1232&$m_\Delta^0$&$S$\ [4]&$\pi_H$ & \cite{17}\\
\label{tab1}
\end{tabular}
\end{table}
\widetext

\begin{table}
\caption{Results for some trinucleon bound state properties.
The computed binding energies are correct within 10 keV only.
Thus, the last digit in rows $E_T$, $\Delta E_2$ and $\Delta E_3$ of this table
are not significant on an absolute scale. The last digit is, however,
significant for relative changes, and this is the reason why it is quoted --
against our practice in other papers. The nucleonic probabilities
$P_{{\cal L}}$ in the wave function are split up according to total
orbital angular momentum ${\cal L}$ and for ${\cal L}=0$ also according to
the symmetry properties of the orbital wave function components in the
standard way.}
\begin{tabular}{lrrrrrrrrr}
&Paris&$H$(1)&$H$(2)&$H$(3)&$H$(4)&$S$(1)&$S$(2)&$S$(3)&$S$(4)\\\tableline
$E_T$[MeV] &-7.381 &-7.849&-7.866&-7.885&-7.912 &-7.627&-7.636&-7.643&-7.667\\
$\Delta E_2$[MeV]&  -    & 0.456  & 0.425 &0.494& 0.460 &0.248 &0.227&0.272
&0.258\\
$\Delta E_3$[MeV]& -& -0.924 &-0.910&-0.998
&-0.991& -0.494 &-0.482&-0.534&-0.544\\
$P_S$[\%]   &90.13  & 88.23 & 88.35 & 88.06
& 88.20 & 89.05 & 89.13 & 88.95 & 89.00 \\
$P_{S'}$[\%]&1.40  & 1.24 & 1.23 & 1.23 & 1.22 & 1.31 & 1.31 & 1.30 & 1.30 \\
$P_P$[\%]   &0.06  &0.08 & 0.08 & 0.09 & 0.08  & 0.08 & 0.08  & 0.08 & 0.08 \\
$P_D$[\%]   &8.41  & 8.68 & 8.69 & 8.70 & 8.71 & 8.61 &  8.61 & 8.62 & 8.63\\
$P_\Delta$[\%]&  -  & 1.71 & 1.64  & 1.93 & 1.79 & 0.92 & 0.87 & 1.05 &1.00 \\
$P_\pi$[\%]  &  -  & 0.06 &  -    & -    &  -    & 0.04 &  -    &  -    &-
\label{tab2}
\end{tabular}
\end{table}
\end{document}